\documentclass[twocolumn,prl,preprintnumbers,nofootinbib]{revtex4}

\usepackage{graphicx}
\usepackage{bm}
\usepackage{epsfig}
\usepackage{slashed}

\newcommand{\lsim}{~{}_{\textstyle\sim}^{\textstyle <}~}

\newcommand{\beq}{\begin{equation}}
\newcommand{\eeq}{\end{equation}}
\newcommand{\bea}{\begin{eqnarray}}
\newcommand{\eea}{\end{eqnarray}}

\begin{document}

\preprint{LAUR-07-3194}

\title{The Standard Model prediction for $R^{(\pi, K)}_{e/\mu} $}
\author{Vincenzo Cirigliano$^1$} 
\author{Ignasi  Rosell$^2$}
\affiliation{
$^1$~Theoretical Division, Los Alamos National Laboratory, Los Alamos NM 87544, USA\\
$^2$~Departamento de Ciencias F\'isicas, Matem\'aticas y de la Computaci\'on,
Universidad CEU Cardenal Herrera, San Bartolom\'e 55, E-46115 Alfara del
Patriarca, Val\`encia,  Spain
} 


\begin{abstract}
We  study  the ratios 
$R_{e/\mu}^{(P)} \equiv 
 \Gamma( P \to e \bar{\nu}_e [\gamma] )/ \Gamma( P \to \mu \bar{\nu}_\mu [\gamma] )$    ($P=\pi,K$)  
in  Chiral Perturbation Theory to order $e^2 p^4$.   
We complement the 
two-loop effective theory results 
with a matching calculation  of the 
counterterm,  
finding  $R_{e/\mu}^{(\pi)} = ( 1.2352 \pm 0.0001 ) \times 10^{-4}$  
and  $R_{e/\mu}^{(K)} = ( 2.477 \pm 0.001 ) \times 10^{-5}$. 
\end{abstract}

\pacs{Valid PACS appear here}
\maketitle

{\it Introduction} -  
The ratio $R_{e/\mu}^{(P)} \equiv  \Gamma( P \to e \bar{\nu}_e [\gamma] )/ 
\Gamma( P \to \mu \bar{\nu}_\mu [\gamma] )$    ($P=\pi,K$) 
is  helicity-suppressed in the Standard Model (SM), 
due to the $V-A$ structure of charged current couplings. 
It is therefore a  sensitive probe of  all  SM extensions 
that induce pseudoscalar currents and  
non-universal corrections to the  lepton couplings~\cite{Bryman:1993gm}, such as  
the minimal supersymmetric SM~\cite{susy-refs}. 
Effects from weak-scale new physics  
are expected in the range  $(\Delta R_{e/\mu})/R_{e/\mu} \sim  10^{-4} - 10^{-2}$  
and there is a realistic chance to detect or constrain them 
because:  (i) ongoing experimental searches plan to reach a fractional uncertainty of 
 $(\Delta R^{(\pi)}_{e/\mu})/R^{(\pi)}_{e/\mu} \lsim  5 \times  10^{-4}$~\cite{exp-new}
and   $(\Delta R^{(K)}_{e/\mu})/R^{(K)}_{e/\mu} \lsim 3 \times 10^{-3}$~\cite{exp-new-K}, 
which represent respectively a factor of  $5$  and $10$ improvement
over current errors~\cite{PDG}. 
(ii) The SM theoretical uncertainty  can be pushed below this level,  since to a first approximation the strong interaction dynamics cancels out in the ratio $R_{e/\mu}$ and hadronic structure dependence  
appears only through electroweak corrections. Indeed, the most recent  
theoretical predictions read  $R^{(\pi)}_{e/\mu} = (1.2352   \pm  0.0005) \times 10^{-4}$~\cite{MS93}, 
$R^{(\pi)}_{e/\mu} = (1.2354   \pm  0.0002) \times 10^{-4}$~\cite{Fink96}, and 
$R^{(K)}_{e/\mu} = (2.472   \pm  0.001) \times 10^{-5}$~\cite{Fink96}.     
The authors of  Ref.~\cite{MS93} provide a general parameterization of the 
hadronic effects and estimate the induced uncertainty  via  dimensional analysis. 
On the other hand, in Ref.~\cite{Fink96}  the hadronic component is calculated by 
modeling the low- and intermediate-momentum region of the loops involving virtual photons.   

With the aim to improve the existing theoretical status,  
we have  analyzed 
$R_{e/\mu}$   within  Chiral Perturbation Theory (ChPT), 
the low-energy effective field theory (EFT) of QCD.  The key feature of this  framework is that it 
provides  a controlled expansion of the amplitudes  in terms of the masses of pseudoscalar mesons and charged leptons  
($p\sim m_{\pi,K, \ell}/\Lambda_\chi$, with $\Lambda_\chi \sim 4 \pi F_\pi \sim 1.2 \,  {\rm GeV}$),  and the electromagnetic  coupling ($e$).   
Electromagnetic corrections to (semi)-leptonic 
decays of   $K$ and $\pi$ 
have been worked out to  $O(e^2 p^2)$~\cite{Knecht:1999ag,Semileptonic}, but 
had never been pushed  to  $O(e^2 p^4)$, as  required for $R_{e/\mu}$. 
In this letter we report the results of our analysis of $R_{e/\mu}$ to $O(e^2 p^4)$, 
deferring the full details to a separate publication~\cite{long}.  
To the order we work,  
$R_{e/\mu}$ features 
both model independent double chiral logarithms (previously neglected) 
and an a priori unknown low-energy  coupling (LEC),  
which we estimate by means of  a matching calculation
in large-$N_C$ QCD.
The inclusion of both effects 
allows us to further 
reduce the  theoretical uncertainty 
and to put its estimate on more solid ground.

Within the chiral power counting,  $R_{e/\mu}$ is  written as:
\bea
R_{e/\mu}^{(P)} &= & R_{e/\mu}^{(0),(P)}   
\, \Bigg[   1 + 
\Delta_{e^2 p^2}^{(P)} +   \Delta_{e^2 p^4}^{(P)}  + \Delta_{e^2 p^6}^{(P)}   +  ...  
\Bigg]  \ \ \ \  \\  
R_{e/\mu}^{(0),(P)} & =& \frac{m_e^2}{m_\mu^2}  \left(  \frac{m_P^2 - m_e^2}{m_P^2 - m_\mu^2} 
 \right)^2  ~. 
\label{eq:R0}
\eea
The leading electromagnetic correction  $\Delta_{e^2 p^2}^{(P)} $
 corresponds to  the point-like approximation for pion and kaon, and its 
 expression is well known~\cite{Kinoshita:1959ha,MS93}.   
Neglecting terms of order $(m_e/m_\rho)^2$,  the most general parameterization of the 
NLO  ChPT contribution 
can be written in the form 
\begin{widetext}
\vspace{-.5cm}
\beq
 \Delta_{e^2 p^4}^{(P)} = \frac{\alpha}{\pi} \frac{m_\mu^2}{m_\rho^2} 
\left(c_2^{(P)}  \, \log \frac{m_\rho^2}{m_\mu^2}  
+  c_3^{(P)} 
+ c_4^{(P)} (m_\mu/m_P) \right) + 
\frac{\alpha}{\pi}  \frac{m_P^2}{m_\rho^2} \,  \tilde{c}_{2}^{(P)}  \, \log \frac{m_\mu^2}{m_e^2} ~ , 
\label{eq:dele2p4}
\eeq
\vspace{-0.2cm}
\end{widetext}
which highlights the dependence on lepton masses. 
The dimensionless constants  $c_{2,3}^{(P)}$  do not  depend on the lepton mass 
but depend logarithmically on hadronic masses, while  $c_4^{(P)} (m_\mu/m_P) \to 0$ 
as $m_\mu \to 0$. 
(Note that  our $c_{2,3}^{(\pi)}$  do not coincide with 
$C_{2,3}$ of Ref.~\cite{MS93}, because their $C_{3}$
is not constrained to be $m_\ell$-independent.)
Finally,  depending 
on the treatment of real photon emission,  one has to include in  $R_{e/\mu}$ 
terms arising from the 
structure dependent contribution to  $P \to e \bar{\nu}_e \gamma$~\cite{Bijnens:1992en}, 
that are formally of $O(e^2 p^6)$, but are not helicity suppressed 
and behave as $\Delta_{e^2 p^6} \sim   \alpha/\pi  \,  (m_P/m_\rho)^4  \, (m_P/m_e)^2$.  

\begin{figure}[t!]
\includegraphics[width=8.0cm]{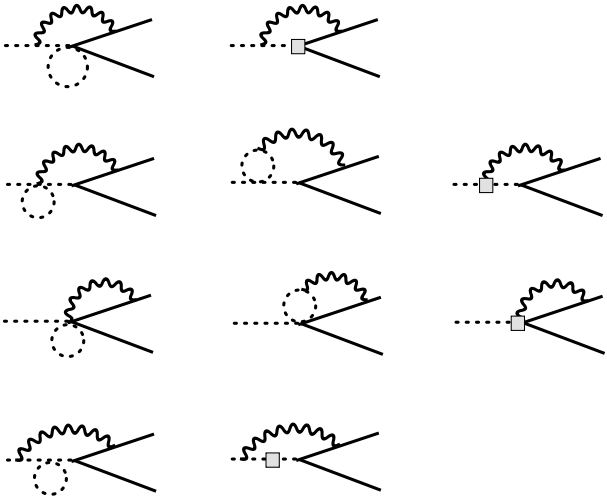}
\caption{One- and two-loop 1PI topologies contributing to $R_{e/\mu}$ to order $e^2 p^4$. 
Dashed lines represent pseudoscalar mesons, solid lines fermions an wavy lines photons. 
Shaded squares indicate vertices from the $O(p^4)$ effective lagrangian. 
}
\label{fig:fig1}
\end{figure}

{\it The calculation} -  
In order to calculate the various coefficients $c_i^{(P)}$ within 
ChPT to $O(e^2 p^4)$,   one has to consider (i)  two-loop graphs 
with vertices from the lowest order effective lagrangian ($O(p^2)$);  
(ii) one-loop graphs with one insertion from the NLO lagrangian~\cite{GL851}  
($O(p^4)$); 
(iii) tree-level diagrams with  insertion of  a local counterterm of 
$O(e^2 p^4)$. 
In Fig.~\ref{fig:fig1}  we show all the  relevant one- and two-loop 1PI  topologies  
contributing to $R_{e/\mu}$. Note that all diagrams 
in which the virtual photon does not connect to the charged lepton line 
have a trivial dependence on the lepton 
mass and  drop when taking the ratio of $e$ and $\mu$ rates.  
We work in Feynman  gauge and use  dimensional regularization to deal with 
ultraviolet (UV)  divergences.

By suitably grouping the 1PI  graphs of  Fig.~\ref{fig:fig1} with external leg corrections, 
it is possible to show~\cite{long} that the effect of the  $O(e^2p^4)$ diagrams amounts to:
(i) a renormalization of the meson mass $m_P$ and decay constant $F_P$ in the 
one-loop result $\Delta_{e^2 p^2}^{(P)}$; 
(ii) a genuine shift to the invariant amplitude $T_\ell \equiv  
T(P^+ (p)  \to \ell^+ (p_\ell) \nu_{\ell} (p_\nu))$. 
This correction can be expressed  as the  convolution of a known kernel with the
vertex function 
$ 
{\cal T}_{\mu \nu} =  
 1 /( \sqrt{2} F)  
 \int d x \ e^{i q x + i W y} \   \langle 0  | T (J^{EM}_{\mu}  (x)  \, (V_\nu - A_\nu) (y)  | \pi^+ (p)  \rangle $
(with $V_\mu (A_\mu) = \bar{u} \gamma_\mu (\gamma_5) d$),   
once the Born term has  been subtracted from the latter. 
Explicitly,  in the case of pion decay one has ($W=p-q$, $\epsilon_{0123}  = + 1$)    
\begin{widetext}
\vspace{-.5cm}
\bea
\delta T_\ell^{e^2 p^4}  & = & 2 G_F V_{ud}^* e^2 F  \   \int  \frac{d^d q}{(2 \pi)^d} \  
\frac{ \bar{u}_L (p_\nu) \gamma^\nu \left[  - (\slashed{p}_{\ell} - \slashed{q})  + m_{\ell}  \right] \gamma^\mu  v(p_\ell)}{
\left[q^2 - 2 q \cdot p_{\ell} + i \epsilon \right] \left[q^2 - m_\gamma^2  + i \epsilon \right ]} \ \,
 {\cal T}_{\mu \nu} (p,q) 
\label{eq:convolution1}
\\
{\cal T}^{\mu \nu} (p,q) &=&  i V_1 (q^2, W^2)   \, \epsilon^{\mu \nu \alpha \beta} q_\alpha p_\beta  
- A_1 (q^2, W^2) \,  \left(  q \cdot p  g^{\mu \nu} - p^\mu q^\nu \right) 
- (A_2(q^2, W^2)  - A_1(q^2,  W^2) )  \left( q^2 g^{\mu \nu}  - q^\mu q^\nu \right) 
\nonumber \\
&+& 
 \left[
\frac{(2 p - q)^\mu (p - q)^\nu}{2 p \cdot q - q^2}   - 
\frac{q^\mu (p - q)^\nu}{q^2}  
\right]  \, \left( F_V^{\pi \pi} (q^2)  - 1\right)   ~ . 
\label{eq:correlator2}
\eea
\end{widetext}
To the order we work,  the form factors $V_1(q^2, W^2)$,  $A_i (q^2, W^2)$  and $F_V^{\pi \pi}(q^2)$  have to be evaluated to $O(p^4)$ in ChPT in $d$-dimensions. 
Their expressions are well known for $d=4$~\cite{Bijnens:1992en}  and have been generalized to any $d$~\cite{long}. 
So the relevant $O(e^2 p^4)$ amplitude is obtained by calculating a set of 
one-loop diagrams with effective 
local ($V_1$ and $A_1$) and non-local   ($A_2$ and $F_V^{\pi \pi}$) $O(p^4)$ vertices. 
The final result can be expressed in terms of one-dimensional integrals~\cite{long}. 

While $c_{2,4}^{(P)}$ and $\tilde{c}_{2}^{(P)}$  are parameter-free predictions of ChPT (they 
depend only on $m_{\pi,K}$,  $F_\pi$, and the LECs
$L_{9,10}$  determined in other processes~\cite{GL851}), 
$c_3^{(P)}$ contains an ultraviolet (UV) divergence, indicating the need to introduce in 
the effective theory a local operator of $O(e^2 p^4)$, with an associated LEC.  
The physical origin of the UV divergence is clear: 
when  calculating  $ \delta T_\ell^{e^2 p^4}$ in the EFT approach, 
we use the $O(p^4)$  ChPT representation of the form factors appearing  
in Eq.~\ref{eq:correlator2} (${\cal T}_{\mu \nu} \to {\cal T}_{\mu \nu}^{\rm ChPT}$).     
 While this representation is valid at scales below $m_\rho$
(and generates the correct single- and double-logs upon integration in $d^d q$) 
it leads to the incorrect UV behavior of the integrand in Eq.~\ref{eq:convolution1}, 
which is instead dictated by the Operator Product Expansion (OPE) for the 
$\langle V V P \rangle$ and  $\langle V A P \rangle$ correlators.  
So  in order to estimate the finite local contribution (dominated by the UV region)
we need a QCD representation of the correlators valid for momenta 
beyond the chiral regime (${\cal T}_{\mu \nu} \to {\cal T}_{\mu \nu}^{\rm QCD}$) .     
This program is  feasible only  within 
an approximation scheme to QCD.  We have used a truncated 
version of large-$N_C$  QCD,  in which 
the correlators are approximated by meromorphic functions, 
representing 
the exchange of a {\it finite}  number  of narrow resonances, 
whose couplings are fixed by requiring that the 
vertex functions $\langle \pi | V A | 0  \rangle$ and  $\langle \pi | V V| 0  \rangle$
obey the leading and next-to-leading OPE behavior at large $q$~\cite{largeN}. 
This procedure allows us to obtain a simple analytic form for the local coupling (see  Eq.~\ref{eq:CT}).  

{\it  Results}  -  The  results for $c_{2,3,4}^{(P)}$ and $\tilde{c}_{2}^{(P)}$ depend on the definition of the inclusive 
rate  $\Gamma (P \to \ell \bar{\nu}_\ell [\gamma])$.  
The radiative  amplitude is the sum of the  inner bremsstrahlung component  ($T_{IB}$)
of $O(e p)$ and a structure 
dependent  component ($T_{SD}$) of $O(e p^3)$~\cite{Bijnens:1992en}.   
The experimental  definition of $R_{e/\mu}^{(\pi)}$ is fully inclusive on the radiative mode, so 
that $\Delta_{e^2 p^4}^{(\pi)}$ receives a contribution  from the interference
of $T_{IB}$ and $T_{SD}$, 
and one also has to include the effect of $\Delta_{e^2 p^6}^{(\pi)} \propto |T_{SD}|^2$.   
The usual experimental definition of  $R_{e/\mu}^{(K)}$ 
corresponds to including the effect of 
$T_{IB}$ in $\Delta_{e^2 p^2}^{(K)}$ (dominated by soft photons) and excluding altogether 
the effect of $T_{SD}$: consequently $c_n^{(\pi)} \neq c_n^{(K)}$. 
%
%

{\it Results for $R_{e/\mu}^{(\pi)}$} - Defining $\bar{L}_9 \equiv (4 \pi)^2  L_{9}^r (\mu)$, 
$\ell_P \equiv \log (m_P^2/\mu^2)$  ($\mu$ is the chiral renormalization scale),   
$\gamma \equiv  A_1(0,0) / V_1 (0,0)$,  $z_\ell \equiv (m_\ell/m_\pi)^2$, we find:   
\begin{widetext}   
\vspace{-.5cm}
\bea
c_2^{(\pi)} &=& 
\frac{2}{3} \, m_\rho^2  \, \langle r^2 \rangle_{V}^{(\pi)}
+   3 \, \left( 1 - \gamma \right)   \, \frac{m_\rho^2}{(4 \pi F)^2}    \qquad \qquad \qquad 
\tilde{c}_2^{(\pi)}  \,  =  \, 0  
\\
c_3^{(\pi)} &=& -  \frac{m_\rho^2}{(4 \pi F)^2}  
\Bigg[  \frac{31}{24} - \gamma + 4 \, \bar{L}_9   + 
\left(\frac{23}{36}  - 2 \, \bar{L}_9  + \frac{1}{12} \ell_K \right) \ell_\pi  + \frac{5}{12} \ell_\pi^2 
 + \frac{5}{18} \ell_K +  \frac{1}{8} \ell_K^2  
\nonumber \\
&+& \left( \frac{5}{3} - \frac{2}{3} \gamma \right) \, \log \frac{m_\rho^2}{m_\pi^2} 
+ \left( 2 + 2 \,  \kappa^{(\pi)} - \frac{7}{3} \gamma  \right) \, \log \frac{m_\rho^2}{\mu^2} 
+  K^{(\pi)} (0)   \Bigg]  \ + \ c_{3}^{CT} (\mu) 
\label{eq:C3pi}
\\
c_4^{(\pi)} (m_\ell) &=& 
-  \frac{m_\rho^2}{(4 \pi F)^2}   \left\{ 
\frac{z_\ell}{ 3  (1 - z_\ell)^2} 
\left[    \Big( 4 ( 1 - z_\ell)   + ( 9 - 5 z_\ell )  \log z_\ell  \Big)
+ 2 \,   \gamma  \,  \Big( 1 - z_\ell  + z_\ell \log z_\ell \Big)    \right] 
\right. 
\nonumber \\
&+& \left.  \left( \kappa^{(\pi)}  + \frac{1}{3} \right) \frac{  \, z_\ell}{2 ( 1 - z_\ell)} \, \log z_\ell   
+ K^{(\pi)}  (z_\ell) - K^{(\pi)} (0)   \right\} 
\label{eq:C4pi}
\eea
\end{widetext}    
where $\kappa^{(\pi)}$ is related to the $O(p^4)$ pion charge radius by:  
\beq
\kappa^{(\pi)} \equiv  4 \, \bar{L}_9 - \frac{1}{6} \ell_K - \frac{1}{3} \ell_\pi - \frac{1}{2} 
=   \frac{(4 \pi F)^2} {3}  \, \langle r^2 \rangle_{V}^{(\pi)} ~.
\label{eq:klog}
\eeq
The function $K^{(\pi)} (z_\ell)$, whose expression will be given in Ref.~\cite{long},  
does not contain any large logarithms and gives a small 
fractional contribution to $c_{3,4}^{(\pi)}$.  

As anticipated, $c_2^{(\pi)}$  is a parameter-free prediction of ChPT. 
Moreover,  we find $\tilde{c}_2^{(\pi)}=0$, as expected due  to a cancellation of real- and virtual-photon 
effects~\cite{Marciano:1976jc}.
Finally,  $c_{3}^{(\pi)}$ encodes calculable chiral corrections (as does  $c_4 (m_\ell)$) 
and a local counterterm  $c_3^{CT} (\mu)$, for which our matching procedure~\cite{long} gives
($z_A \equiv  m_{a_1}/m_\rho$):
\bea
c_3^{CT} (\mu) \! &=& \!
-  \frac{19 \, m_\rho^2}{9 (4 \pi F)^2}   +  \left(  \!
\frac{ 4 \, m_\rho^2}{3 (4 \pi F)^2}  +  \frac{7 + 11 z_A^2}{6      z_A^2}  \! 
\right)   \log\frac{m_\rho ^2 }{\mu^2}  
\nonumber \\
 &+& \frac{ 37 - 31 z_A^2   + 17 z_A^4 - 11 z_A^6}{ 36 z_A^2  ( 1 - z_A^2)^2}
 \nonumber \\
&-&  \frac{ 7 - 5 z_A^2   - z_A^4  + z_A^6}{ 3 z_A^2  ( - 1 + z_A^2)^3} \, \log z_A ~.
\label{eq:CT}
\eea
Numerically,  using  $z_A = \sqrt{2}$, we find $c_3^{CT} (m_\rho) = -1.61$,   
implying that the counterterm induces a sub-leading correction to $c_3$ (see Table~\ref{tab:tab1}). 
The scale dependence of 
$c_{3}^{CT}(\mu)$ 
partially cancels the scale dependence 
of the  chiral loops (our procedure captures all the "single-log" scale dependence). 
Taking a very conservative attitude we assign to $c_3$  an uncertainty equal to 
$100\%$  of the local contribution ($| \Delta c_3 | \sim 1.6$)  plus the effect of residual 
renormalization scale dependence, obtained by varying the scale $\mu$ in the range 
$0.5 \to 1$ GeV   ($| \Delta c_3|  \sim 0.7$), leading to 
$\Delta c_3^{(\pi,K)} = \pm 2.3$. 
Full numerical values of $c_{2,3,4}^{(\pi)}$ are reported in Table~\ref{tab:tab1}, with 
uncertainties due to matching  procedure 
 and input parameters ($L_9$ and $\gamma$~\cite{pocanic}).

As a check on our calculation, we have verified that if we neglect $c_{3}^{CT}$ and pure two-loop effects,  and if we use $L_9 = F^2/(2 m_\rho^2)$  (vector meson dominance), 
our results for $c_{2,3,4}^{(\pi)}$ are   
fully consistent with previous analyses of the leading 
structure dependent corrections based on current algebra~\cite{terentev,MS93}.    
Moreover, our numerical value of  
$\Delta_{e^2 p^4}^{(\pi)}$ reported in  Table~\ref{tab:tab2} is very close to the corresponding 
result  in Ref.~\cite{MS93}, 
$\Delta_{e^2 p^4}^{(\pi)} =  (0.054 \pm 0.044) \times 10^{-2}$. 

For completeness we report here the contribution to $\Delta_{e^2 p^6}^{(\pi)}$  induced by structure dependent 
radiation: 
\bea
\Delta_{e^2 p^6}^{(\pi)}  &=&  
\frac{\alpha}{2  \pi}  \, \frac{m_\pi^4}{(4 \pi F)^4} \left(1 + \gamma^2 \right)  \ 
\Big[ 
\frac{1}{30 \, z_e} - \frac{11}{60} + \frac{z_e}{20 (1 - z_e)^2}  
\nonumber \\
&\times& 
\left( 
12 - 3 z_e - 10 z_e^2 + z_e^3 + 20\,  z_e \log z_e 
\right)
\Big]~.
\eea

{\it Results for $R_{e/\mu}^{(K)}$} -  
In this case  we have: 
\bea
c_2^{(K)} &=&
\frac{2}{3} \, m_\rho^2  \, \langle r^2 \rangle_{V}^{(K)}
+ \frac{4}{3} \,  \left( 1 - \frac{7}{4} 
\gamma \right)\, \frac{m_\rho^2}{(4 \pi F)^2}   
\\
\tilde{c}_2^{(K)}  &  =  &  \frac{1}{3} \left(1 - \gamma \right)  \,
 \frac{m_\rho^2}{(4 \pi F)^2}   
\eea
where $\langle r^2 \rangle_{V}^{(K)}$
is the $O(p^4)$ kaon charge radius.   
$c_3^{(K)}$ is obtained from $c_3^{(\pi)}$ by replacing $31/24 - \gamma \to $ 
$- 7/72  - 13/9 \, \gamma$, by dropping the term proportional to 
$\log m_\rho^2/m_\pi^2$, 
and by inter-changing everywhere else the label $\pi$ with $K$ (masses, 
$\ell_\pi \to \ell_K$, etc.). 
$c_4^{(K)}$ is obtained from $c_4^{(\pi)}$ by keeping only the second line of Eq.~\ref{eq:C4pi} 
and inter-changing the labels $\pi$ and $K$. 
The numerical values of $c_{2,3,4}^{(K)}$ and $\tilde{c}_{2}^{(K)}$   
are reported in Table~\ref{tab:tab1}.

\begin{table}[t!]
\begin{center}
\begin{tabular}{|c|c|c|}
\hline
  & $(P=\pi)$  & $(P=K)$    \\[5pt]
\hline
 $\tilde{c}_2^{(P)}$  &   0  &  $ (7.84 \pm 0.07_\gamma) \times 10^{-2}  $ \\
 $c_2^{(P)}$  & $5.2 \pm 0.4_{L_9} \pm 0.01_\gamma$  &  $4.3 \pm 0.4_{L_9} \pm 0.01_\gamma $ \\
 $c_3^{(P)}$   
&   
$ -10.5  \pm 2.3_{\rm m } \pm 0.53_{L_9} 
$
& 
$ -4.73 \pm 2.3_{ \rm m} \pm 0.28_{L_9}
$ 
\\
 $c_4^{(P)} (m_\mu) $  &
$1.69  \pm 0.07_{L_9} $
&  
$ 0.22 \pm 0.01_{L_9} $ 
\\
\hline
\end{tabular}
\end{center}
\caption{Numerical values of the coefficients  $c_n^{(P)}$ of Eq.~\ref{eq:dele2p4} ($P=\pi,K$).
The uncertainties correspond to the input values  $L_9^r (\mu=m_\rho) = 
(6.9 \pm 0.7) \times 10^{-3} $, 
$\gamma= 0.465 \pm 0.005$~\cite{pocanic}, and to the matching procedure (${\rm m}$), 
affecting only $c_3^{(P)}$.}
\label{tab:tab1}
\end{table}

{\it  Resumming leading logarithms}  -     
At  the level of uncertainty considered,  one needs to include higher order 
long distance corrections  to the leading contribution 
$\Delta_{e^2 p^2} \sim - 3 \alpha/\pi \log m_\mu/m_e \sim - 3.7 \%$.  
The leading logarithms  can be summed via the  renormalization group 
and their effect amounts to  multiplying $R_{e/\mu}^{(P)}$ by~\cite{MS93}
\beq
1 + \Delta_{LL} = 
\displaystyle\frac{\left(1 - \frac{2}{3} \frac{\alpha}{\pi} \log \frac{m_\mu}{m_e} \right)^{9/2}}{1 -
 \frac{3 \alpha}{\pi} \log \frac{m_\mu}{m_e}} = 
1.00055~.
\eeq
%

\begin{table}[t!]
\begin{center}
\begin{tabular}{|c|c|c|}
\hline
 & $(P=\pi)$  & $(P=K)$    \\[5pt]
\hline
 $\Delta_{e^2 p^2}^{(P)} \ \, (\%)  $   &   $-3.929 $ &  $ -3.786 $ \\
$\Delta_{e^2 p^4}^{(P)}  \ \, (\%) $ 
 & $0.053 \pm 0.011$  &  $0.135 \pm 0.011$ \\
$\Delta_{e^2 p^6}^{(P)} \ \,  (\%)  $ 
&   
$ 0.073
$
& 
\\
$\Delta_{LL} \ \ (\%) $   &
$ 0.055 $ 
&  
$ 0.055 $ 
\\
\hline
\end{tabular}
\end{center}
\caption{Numerical summary of various electroweak corrections to $R_{e/\mu}^{(\pi,K)}$. }
\label{tab:tab2}
\end{table}

{\it Conclusions} -   
In Table~\ref{tab:tab2} we summarize the various  corrections to 
$R_{e/\mu}^{(\pi,K)}$, which lead to our final results: 
\bea
R_{e/\mu}^{(\pi)} &=& ( 1.2352 \pm 0.0001 ) \times 10^{-4} 
\label{eq:Rpf} \\ 
R_{e/\mu}^{(K)} &=& ( 2.477 \pm 0.001 ) \times 10^{-5} ~ . 
\label{eq:RKf}
\eea
In the case of $R_{e/\mu}^{(K)}$ we have inflated the nominal uncertainty arising from matching  
by a factor of four,  to account for higher order chiral corrections  of expected  size 
$\Delta_{e^2 p^4} \times m_K^2/(4 \pi F)^2$. 
Our results have 
to be compared with the ones of Refs.~\cite{MS93} and \cite{Fink96} reported in the introduction. 
While  $R_{e/\mu}^{(\pi)}$ is  in good agreement with both previous results,  
there is a discrepancy  in $R_{e/\mu}^{(K)}$ that goes well outside the 
estimated theoretical uncertainties. 
We have traced back  this difference  to  the following problems in  Ref.~\cite{Fink96}:
(i) the leading log correction $\Delta_{LL}$ is included  with the wrong sign (this accounts for half  
of the discrepancy);  (ii)  the NLO virtual correction $\Delta_{e^2 p^4}^{(K)}  =  0.058 \%$ 
is not reliable because the hadronic form factors modeled in Ref.~\cite{Fink96} 
do not satisfy the QCD short-distance behavior. 

In conclusion,  by performing the first ever ChPT calculation to $O(e^2 p^4)$,  
we have improved the reliability of both the central value and the uncertainty 
of the  ratios $R_{e/\mu}^{(\pi,K)}$.  
Our final result for $R_{e/\mu}^{(\pi)}$ is consistent with the previous literature, 
while we find a discrepancy in $R_{e/\mu}^{(K)}$,  which we have traced back to 
inconsistencies in the analysis of Ref.~\cite{Fink96}. 
Our results provide a clean basis to detect or constrain non-standard physics in these 
channels  by  comparison with upcoming measurements.

\acknowledgments
{\bf Acknowledgments}  --  We wish to thank M.~ Ramsey-Musolf for collaboration at  an early 
stage of this work,  D.~Pocanic and M.~Bychkov for correspondence on the 
experimental input on $\gamma$, and W.~Marciano and A.~Sirlin for  cross-checks on parts 
of our calculation. This work has been supported in part by the EU MRTN-CT-2006-035482
(FLAVIAnet), by MEC (Spain) under grant FPA2004-00996 and by Generalitat
Valenciana under grant GVACOMP2007-156

\end{document}